\documentclass{PoS}
\usepackage{epsfig}
\usepackage{amsfonts}
\usepackage{amssymb}
\usepackage{amsmath}
\usepackage{simplewick}
\usepackage{graphicx}

\newcommand{\be}{\begin{equation}}
\newcommand{\ee}{\end{equation}}
\newcommand{\bea}{\begin{eqnarray}}
\newcommand{\eea}{\end{eqnarray}}
\newcommand{\tr}{\,\hbox{\rm Tr}}
\newcommand{\dmu}{\hat{\mu}}
\newcommand{\dnu}{\hat{\nu}}

\newcommand{\Vd}{V^\dagger}

\newcommand{\wS}{\ensuremath{\tilde\rho} }

\newcommand{\msbar}{{\rm \overline{MS\kern-0.05em}\kern0.05em}}

\title{Recent progress on chiral symmetry breaking in QCD}

\ShortTitle{Recent progress on chiral symmetry breaking in QCD}

\author{\speaker{Leonardo Giusti}\\
Dipartimento di Fisica, Universit\`a di Milano Bicocca\\
and INFN, Sezione di Milano Bicocca,\\
Piazza della Scienza 3, I-20126 Milano, Italy\\
E-mail: \email{Leonardo.Giusti@mib.infn.it}}

\abstract{I review recent progress achieved on the lattice in the 
quantitative comprehension of chiral symmetry breaking in QCD. Emphasis
is given to the recent precise computations of the spectral density of the
Dirac operator in the continuum limit, and of the topological
susceptibility.}

\FullConference{The 33rd International Symposium on Lattice Field Theory\\
		14 -18 July 2015\\
		Kobe International Conference Center, Kobe, Japan*}

\begin{document}

\section{Introduction}
There is overwhelming evidence from the lattice that the chiral symmetry group
SU$(N_f)_L\times$ SU$(N_f)_R$ of the action of Quantum Chromodynamics (QCD) with
a small number $N_f$ of massless flavors breaks spontaneously to SU$(N_f)_{L+R}$.
This progress became possible thanks to the steady increase of the
computer power made available to our community, and to the impressive algorithmic and
technical progress achieved over the last decade in the numerical simulation of
lattice QCD with light dynamical
fermions~\cite{Hasenbusch:2001ne,Luscher:2005rx,Urbach:2005ji,Luscher:2007es,Luscher:2012av}.

The Gell-Mann-Oakes-Renner (GMOR) relation was beautifully observed
in the very first computations of this new generation
of simulations  already at finite lattice
spacing~\cite{Luscher:2005mv,Giusti:2007hk,DelDebbio:2006cn,Boucaud:2007uk}.
Lattice results for ratios of low-lying eigenvalues of the Dirac operator in the
$\epsilon$ regime turned out to be in agreement with the parameter-free
predictions of leading order (LO) Chiral perturbation theory (ChPT) and random matrix
theory~\cite{Fukaya:2007fb}. By now it is standard practice to assume the presence of
spontaneous symmetry breaking in QCD, and fit phenomenologically interesting observables in the
quark mass by applying the predictions of chiral perturbation theory (ChPT)
\cite{Weinberg:1978kz,Gasser:1983yg,Gasser:1984gg}. After only 10 years from the first
simulations with light quarks, we have many results in the $N_f=2$, $N_f=2+1$ and $N_f=2+1+1$
theories with light quarks down to the physical point, or lattice spacings as small
as $~0.05$~fm \cite{Engel:2010my,Fritzsch:2012wq,Durr:2013goa,Blum:2014tka,Bruno:2014jqa}.
There are many determinations of the QCD LO and next-to-leading order (NLO) low-energy constants
(LECs) obtained by comparing the predictions of ChPT with lattice results,
see Ref.~\cite{Aoki:2013ldr} for a comprehensive review\footnote{New (preliminary) determinations of 
some of the LECs of the SU$(2)$ and SU$(3)$ chiral effective theories of QCD with $N_f=2+1$ flavours
which have reported at this conference \cite{Mawhinney:2015sfj,Mawhinney:2015???} quote errors significantly smaller than
in Ref.~\cite{Aoki:2013ldr}.}.

Over the last decade we have also accumulated stronger and stronger evidence that the
breaking due to the quantum anomaly of the U$(1)_L\times$ U$(1)_R$ chiral group to
U$(1)_{L+R}$  is driven by the Witten--Veneziano mechanism~\cite{'tHooft:1976up,Witten:1979vv,Veneziano:1979ec}.
After $10$--$15$ years of exploratory computations with cooling techniques,
well summarized in Ref.~\cite{Vicari:2008jw}, a theoretically well defined
definition of the cumulants of the topological charge was
found~~\cite{Neuberger:1997fp,Hasenfratz:1998ri,Luscher:1998pqa,Giusti:2001xh,
Giusti:2004qd,Luscher:2004fu}. The value of the topological susceptibility
obtained with the Neuberger definition of the charge~\cite{DelDebbio:2004ns}
indeed supports the Witten--Veneziano explanation for the large experimental value
of the $\eta'$ mass. To properly verify the mechanism, however, precise computations at large
$N_c$ with and without fermions are still required.

In the past two years or so, there have been further substantial conceptual, technical and numerical
progress in our quantitative understanding of chiral symmetry breaking in QCD. It is the
aim of this talk to review these advances with particular emphasis on the recent precise
computations of the spectral density of the Dirac operator in the continuum limit, and
of the cumulants of the topological charge distribution with and without fermions.

\section{Spectral density of the Dirac operator}
The spectral density of the Euclidean massless Dirac operator $D$ is defined
as~\cite{Banks:1979yr,Leutwyler:1992yt,Shuryak:1992pi} 
\be
  \rho(\lambda,m)=\frac{1}{V}\sum_{k=1}^{\infty}
  \left\langle\delta(\lambda-\lambda_k)\right\rangle\; ,
\ee
where $i\lambda_1$, $i\lambda_2$, $\ldots$ are its  
(purely imaginary) eigenvalues, the bracket 
$\langle\ldots\rangle$ denotes the QCD expectation
value, $V$ is the volume of the system, and $m$ is the quark mass.
In QCD the density $\rho(\lambda,m)$ is a renormalizable quantity which is unambiguously
defined once the free parameters in the action (coupling constant and quark masses)
have been renormalized~\cite{Giusti:2008vb,DelDebbio:2005qa}.
The Banks--Casher relation~\cite{Banks:1979yr}
\be
   \lim_{\lambda \to 0}\lim_{m \to 0}\lim_{V \to \infty}\rho(\lambda,m)
   =\frac{\Sigma}{\pi}
\ee
links the density at the origin to the chiral condensate
\be\label{eq:Sigrho} 
  \Sigma=-\frac{1}{2}\lim_{m \to 0}\lim_{V \to \infty}
\left\langle \bar\psi \psi\right\rangle\; ,
\ee
where $\psi$ is the quark doublet. It can be read in either directions:
a non-zero spectral density at the origin implies that the symmetry is broken by a
non-vanishing $\Sigma$ and vice versa. The mode number of the Dirac
operator~\cite{Giusti:2008vb}
\be
  \nu(\Lambda,m)=V \int_{-\Lambda}^{\Lambda} {\rm d}\lambda\,\rho(\lambda,m),
\ee  
turns out to be a renormalization-group invariant quantity as it stands.
In presence of a non-zero chiral condensate the modes condense near the
origin, and the mode number grows linearly with $\Lambda$.  It can also
be written as the average number of eigenmodes
of the massive Hermitean operator $D^{\dagger}D+m^2$ with eigenvalues
$\alpha\leq M^2 = \Lambda^2+m^2$. Its (normalized) discrete derivative 
\be\label{eq:discD}
{\tilde\rho}(\Lambda_1,\Lambda_2,m)  =  
\frac{\pi}{2 V} \frac{\nu(\Lambda_2)-\nu(\Lambda_1)}
{\Lambda_2 - \Lambda_1}\;  
\ee
carries the same information as $\rho(\lambda,m)$, but this {\it effective
spectral density} is a more convenient quantity to consider in practice
on the lattice. 

\subsection{Mode number on the lattice}
On a discretized space-time and Wilson-type fermions, it is appropriate 
to define the mode number directly as the average number of eigenmodes
of the squared massive Hermitean Wilson-Dirac operator $D^\dagger_m D_m$
with eigenvalues $\alpha\leq M^2$. In the continuum limit this definition 
converges to the universal one~\cite{Giusti:2008vb}, with a rate
proportional to $a^2$ if non-perturbative O$(a)$-improved Wilson
or Wilson twisted-mass fermions are implemented~\cite{Giusti:2008vb,Cichy:2014yca}.  
Since with those fermions chiral symmetry is explicitly broken at finite lattice
spacing, the spectrum of the Wilson--Dirac 
operator near threshold ($\Lambda=0$) is not protected from large discretization 
effects~\cite{DelDebbio:2005qa,Damgaard:2010cz,Splittorff:2012gp}.
While this region may be of interest for studying the peculiar 
details of those fermions, it is easier to extract universal information 
about the continuum theory  far away from it. This is one of the reasons
for considering on the lattice the effective spectral density in
Eq.~(\ref{eq:discD}). 

\subsection{Mode number in ChPT\label{ssec:ChPT}}
When chiral symmetry is spontaneously broken, the mode number
can be computed in the chiral effective theory. At the NLO it
reads~\cite{Smilga:1993in,Damgaard:1998xy,Giusti:2008vb,Damgaard:2008zs,Damgaard:2010cz,Necco:2011vx} 
\be\label{eq:RNLO}
\nu^{\rm nlo}(\Lambda,m) =  \frac{2 \Sigma \Lambda V}{\pi}  \Big\{ 1 + 
\frac{m \Sigma}{(4\pi)^2 F^4}\Big[3\, \bar l_6 + 1 - \ln(2) 
- 3 \ln\Big(\frac{\Sigma m}{F^2 \bar\mu^2}\Big) + 
f_\nu\left(\frac{\Lambda}{m}\right)\Big]\Big\}\; , 
\ee
where 
\be
f_\nu(x) = x \left[{\rm arctan}(x) - \frac{\pi}{2}\right]
- \frac{1}{x} {\rm arctan}(x) - \ln(x) - \ln(1+x^2)\; .
\ee
The constants $F$ and $\bar l_6$ are, respectively, the pion decay
constant in the chiral limit and a SU$(3|1)$ low-energy effective coupling
renormalized  at the scale $\bar\mu$. The formula in Eq.~(\ref{eq:RNLO}) has 
some interesting properties:
\begin{itemize}
\item for $x \rightarrow \infty$
\be
f_\nu(x) \xrightarrow{x\to\, \infty} -3 \ln(x)\;,
\ee
and therefore at fixed $\Lambda$ the mode number has {\it no chiral 
logarithms} at the NLO when $m \rightarrow 0$;
\item since in the continuum the operator $D^\dagger_m D_m$ has a threshold 
at $\alpha=m^2$, the mode number must satisfy  
\be
\lim_{\Lambda \rightarrow 0} \nu^{\rm nlo}(\Lambda,m) = 0\; ,
\ee
a property which is inherited by the NLO ChPT formula;
\item in the chiral limit $\nu^{\rm nlo}(\Lambda,m)/\Lambda$ 
{\it becomes independent on} $\Lambda$. This is an accident of the 
$N_f=2$ ChPT theory at NLO \cite{Smilga:1993in};
\item the $\Lambda$-dependence in the square brackets on the r.h.s. 
of Eq.~(\ref{eq:RNLO}) is parameter-free. Since 
$\displaystyle m \Sigma/(4\pi F^2)^2>0$, the behaviour of the 
function $f_\nu(x)$ implies that $\nu^{\rm nlo}(\Lambda,m)/\Lambda$ is a 
decreasing function of (small) $\Lambda$ at fixed (small) $m$. 
\end{itemize}
At the NLO the effective spectral density ${\wS}^{\rm nlo}$ 
inherits the same special properties 
of $\nu^{\rm nlo}(\Lambda,m)/\Lambda$: at fixed 
$\Lambda_{1}$ and $\Lambda_{2}$ it has no chiral logarithms when 
$m \rightarrow 0$, it is independent from $\Lambda_{1}$ and 
$\Lambda_{2}$ in the chiral limit, and at non-zero quark mass it is a decreasing 
parameter-free function of~\footnote{It is very weakly
dependent on $(\Lambda_{1}-\Lambda_{2})$ for pairs of values $\Lambda_{1}$ and $\Lambda_{2}$
that we will consider.} 
$(\Lambda_{1} + \Lambda_{2})/2$. For light values of the quark masses, i.e. $10$~MeV or so,
the variations are of the order of a few percent in the range
of $\Lambda$'s we are interested in.
These special properties offer non-trivial tests that the values of $\Lambda$
and $m$ chosen in the simulations are in a regime where NLO ChPT can be applied.

\section{The spectral density in QCD Lite}
In the last two years the spectral density of QCD with two
light flavours has been computed on a rich set of lattices with a
statistical accuracy of a few percent~\cite{Cichy:2013gja,Engel:2014cka,Engel:2014eea}.
The two groups opted for different gluonic and fermionic regularizations. The authors
of Ref.~\cite{Cichy:2013gja} implemented the tree-level Symanzik
improved gluon action and the Wilson twisted mass fermion action so to be able to
use the gauge configurations generated by the ETM Collaboration.
They span the parameter ranges\footnote{If not explicitly stated, the scheme- and 
scale-dependent quantities such as $\Sigma$, $m$, $\Lambda$ and 
$\rho$ are renormalized in the $\msbar$ scheme at $\mu=2$~GeV.} $a=0.054$--$0.085$, $m=16$--$47$ MeV,
and $M=50$--$120$~MeV (corresponding to approximatively $\Lambda=40$--$120$~MeV).
The authors of Refs.~\cite{Engel:2014cka,Engel:2014eea} opted for the
standard Wilson gluonic action and the non-perturbatively
${\cal O}(a)$-improved Wilson fermion action so to profit from the
generation of the gauge configurations carried out by the
CLS community\footnote{https://wiki-zeuthen.desy.de/CLS/CLS.}
and the Alpha
collaboration\footnote{https://www-zeuthen.desy.de/alpha/}~\cite{DelDebbio:2007pz,Fritzsch:2012wq}.
The parameter ranges that they span are
$a=0.048$--$0.075$, $m=6$--$37$ MeV, and $\Lambda=20$--$500$~MeV.
\begin{figure}[t!]
\hspace{-0.25cm}\begin{minipage}{0.35\textwidth}
\vspace{-0.75cm}
  
\includegraphics[width=5.75 cm,angle=270]{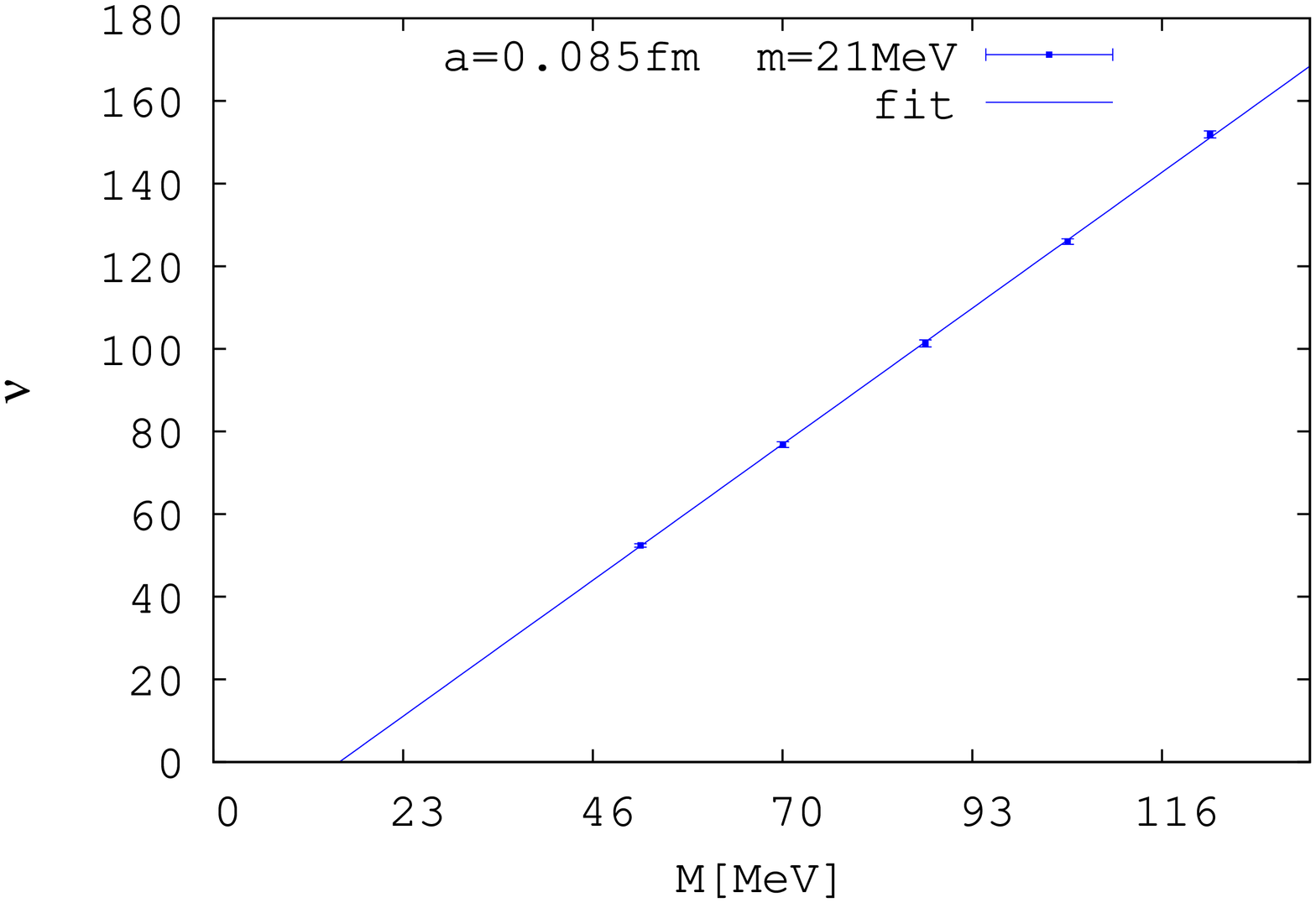}
\end{minipage}
\hspace{15mm}
\begin{minipage}{0.35\textwidth}
\vspace{1.15cm}
  
\includegraphics[width=5.25 cm,angle=0]{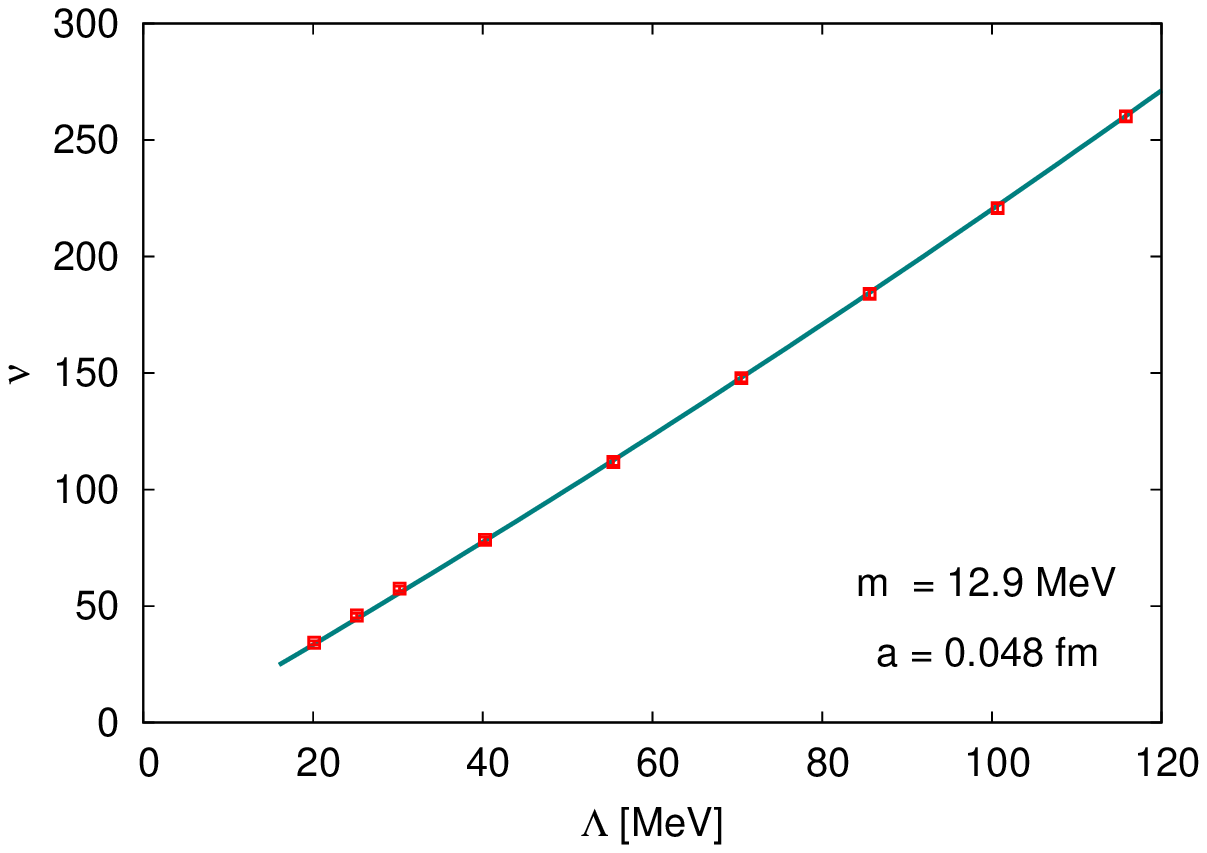}
\end{minipage}
\vspace{-1.00cm}

\caption{Left: the mode number $\nu$ as a function of $M$
for $a=0.085$~fm and $m=21$~MeV from Ref.~\cite{Cichy:2013gja};
the solid line is a linear fit to all 5 points. 
Right: the mode number as a function of $\Lambda$
for $a=0.048$~fm and $m=12.9$~MeV from Refs.~\cite{Engel:2014cka,Engel:2014eea};
a quadratic fit of the data gives  $\nu= -9.0(13) + 2.07(7)\Lambda
+ 0.0022(4)\Lambda^2$. Courtesy of 
Refs.~\cite{Cichy:2013gja,Engel:2014cka,Engel:2014eea}.
}
\label{fig:firstlook}
\end{figure}
In the left plot of Figure~\ref{fig:firstlook} the mode number is shown as a function of $M$
for a lattice with quark mass $m=21$~MeV and
spacing $a=0.085$~fm from Ref.~\cite{Cichy:2013gja}. The right plot in the same figure shows $\nu$ as
a function of $\Lambda$ for $m=12.9$~MeV and $a=0.048$~fm from Ref.~\cite{Engel:2014cka,Engel:2014eea},
a plot which makes manifest that the mode number is a nearly linear function of $\Lambda$ up to
approximatively $100$--$150$~MeV.
The modes do condense near the origin as predicted by the Banks--Casher mechanism. At the percent
precision, however, data show statistically significant deviations from the linear behavior
already below $100$ MeV. Just to guide the eye, a quadratic fit in $\Lambda$ is shown
in the left plot of Figure~\ref{fig:firstlook} where is represented the set of data with lighter
quark masses and which cover a wider range in $\Lambda$. The values of the coefficients, given in the caption,
reveal that the bulk of $\nu$ is given by the linear term, while the constant and the quadratic
term represent ${\cal O}(10\%)$ corrections in the fitted range. This calls for a careful analysis of the
systematics induced by cutoff effects, finite quark mass and finite $\Lambda$ values in order to
reach a precise determination of the spectral density in the continuum and chiral limits at
the origin.

\subsection{Continuum-limit extrapolation}
\begin{figure}[t!]
\hspace{-0.25cm}\begin{minipage}{0.35\textwidth}
  
\includegraphics[width=8.0 cm,angle=0]{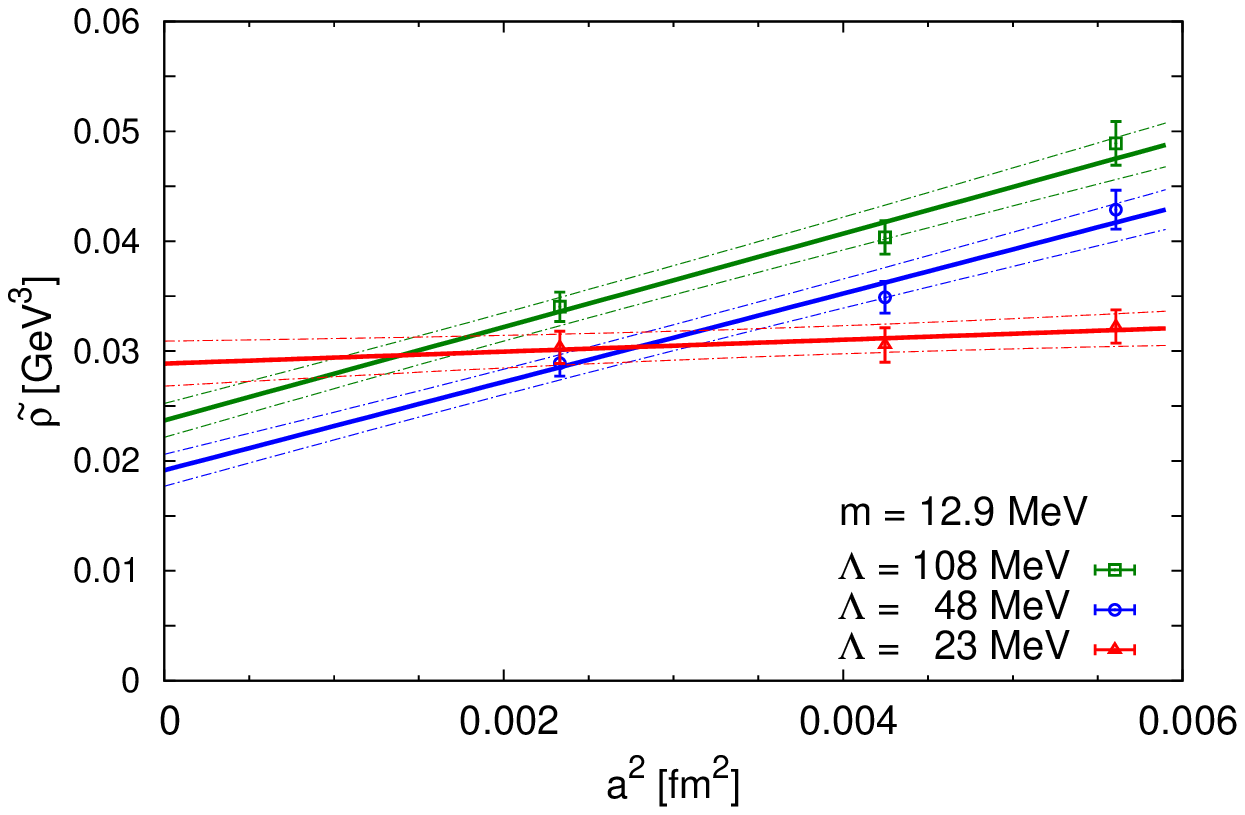}
\end{minipage}
\hspace{25mm}
\begin{minipage}{0.35\textwidth}
  
\includegraphics[width=8.0 cm,angle=0]{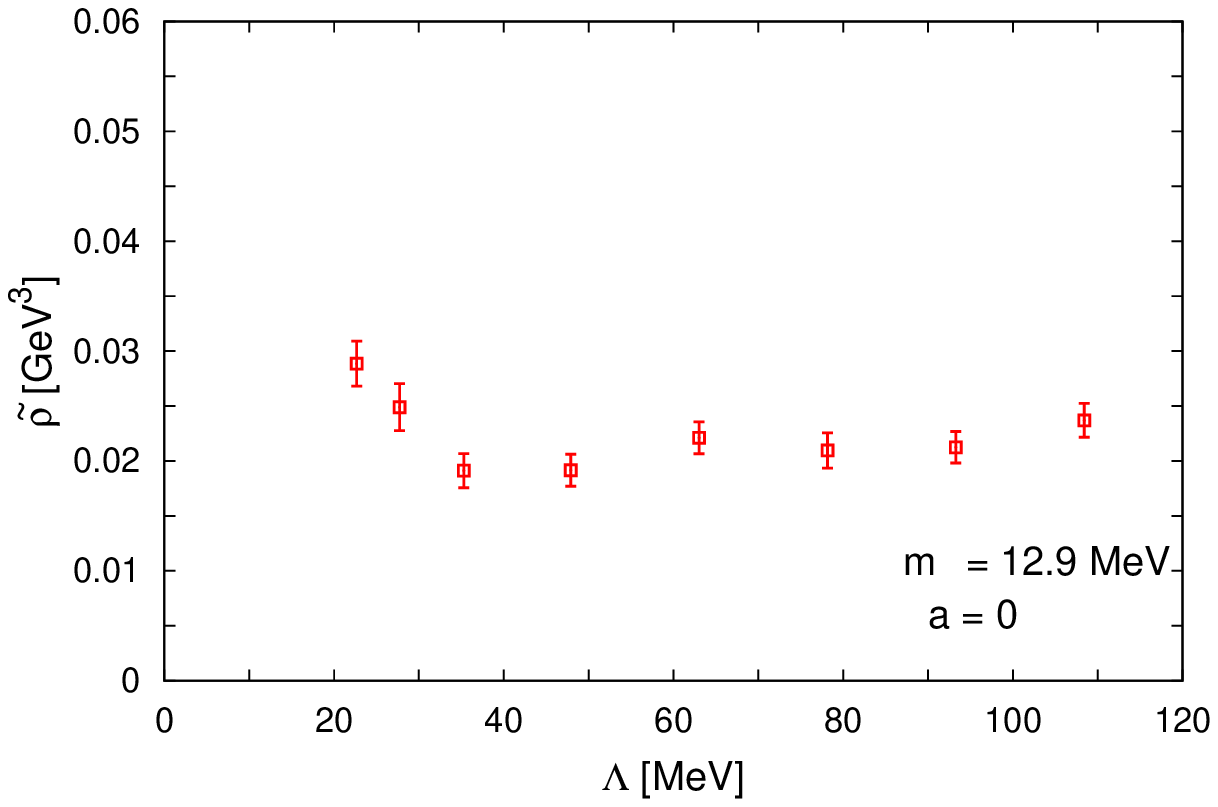}
\end{minipage}
\vspace{-0.25cm}

\caption{Left: continuum limit extrapolation of $\wS$ at the smallest reference quark mass
$m=12.9$~MeV for three values of $\Lambda=(\Lambda_1+\Lambda_2)/2$~\cite{Engel:2014eea}.
  Right: effective spectral density in the continuum limit as a function of $\Lambda=(\Lambda_1+\Lambda_2)/2$
  at the smallest
  reference quark mass $m=12.9$~MeV considered in Ref.~\cite{Engel:2014eea}. Courtesy of 
Refs.~\cite{Engel:2014cka,Engel:2014eea}.
}
\label{fig:conlim}
\end{figure}

Since it is not affected by threshold effects, the effective spectral density
$\wS$ in Eq.~(\ref{eq:discD}) is the primary observable the authors of
Ref.~\cite{Engel:2014cka,Engel:2014eea} focus on. The nearly linear
behaviour of the mode number  manifest itself in the (almost) flatness of ${\wS}$
in the same range.  Because the action and the mode
number are ${\cal O}(a)$-improved, the Symanzik effective theory analysis predicts
that discretization errors start at ${\cal O}(a^2)$. In order to remove them, at every
lattice spacing the authors of Ref.~\cite{Engel:2014cka,Engel:2014eea} matched three quark mass
values ($m=12.9$, $20.9$, $32.0$~MeV) by interpolating $\wS$ linearly in $m$
for each value of $\Lambda$. Within the statistical errors all sets of data are compatible
with a linear dependence in $a^2$, and thus each triplet of points are extrapolated linearly
in $a^2$ to the continuum limit independently, e.g. see left plot in Figure~\ref{fig:conlim}.
The results for $\wS$ at $m=12.9$~MeV in the continuum limit are shown as a function
of $\Lambda=(\Lambda_1+\Lambda_2)/2$ in the right plot of the same Figure. A similar $\Lambda$-dependence
is observed at the two other reference masses. 

It is worth noting that no assumption on the presence of 
spontaneous symmetry breaking was needed so far. These results, however, point to the fact
that the spectral density of the Dirac operator in two-flavour QCD is (almost) constant 
in $\Lambda$ near the origin at small quark masses. This is consistent 
with the expectation from the Banks--Casher relation in the presence of 
spontaneous symmetry breaking. As discussed in section~\ref{ssec:ChPT}, in this case the NLO ChPT 
indeed predicts an almost flat function in (small) $\Lambda$ at (small) finite quark 
masses which is parameter-free once the pion mass and decay constant are measured. 

\subsection{Chiral limit}
\begin{figure}[t!]
\hspace{-0.375cm}\begin{minipage}{0.35\textwidth}
\includegraphics[width=8.0 cm,angle=0]{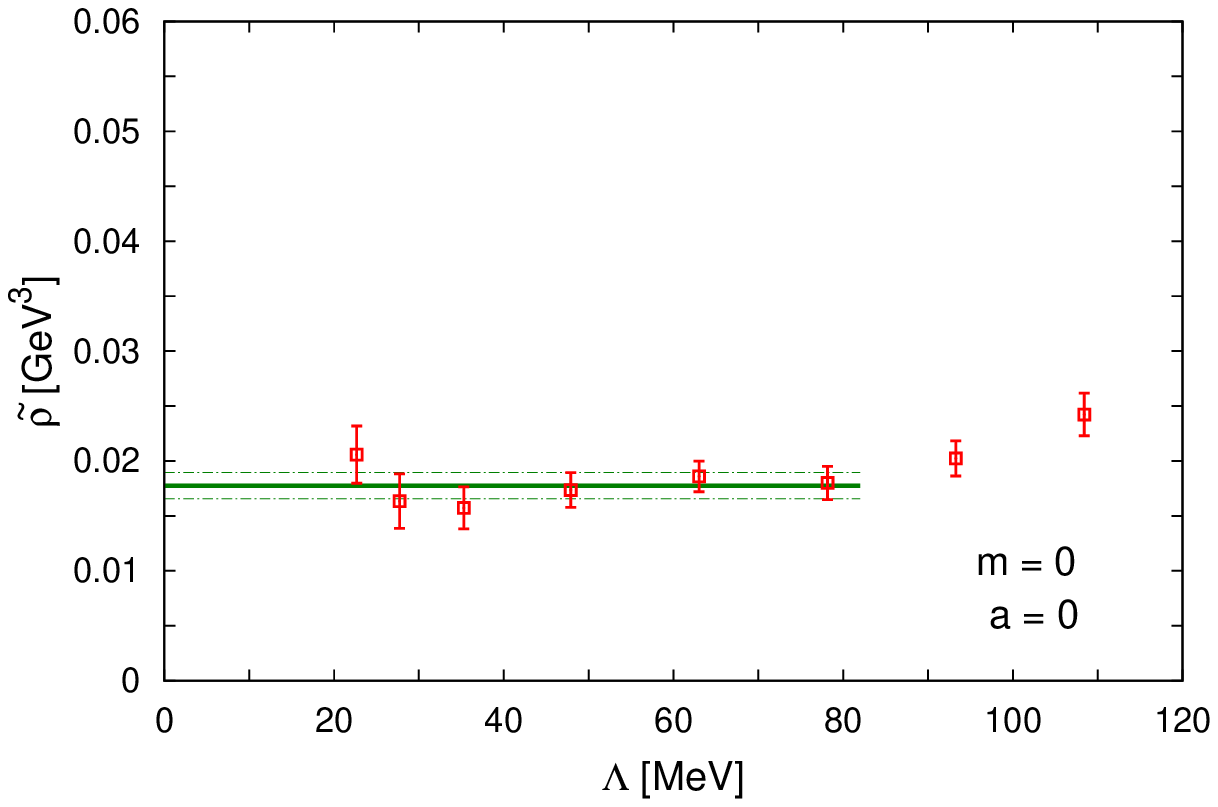}
\end{minipage}
\hspace{25mm}
\begin{minipage}{0.35\textwidth}
\includegraphics[width=8.0 cm,angle=0]{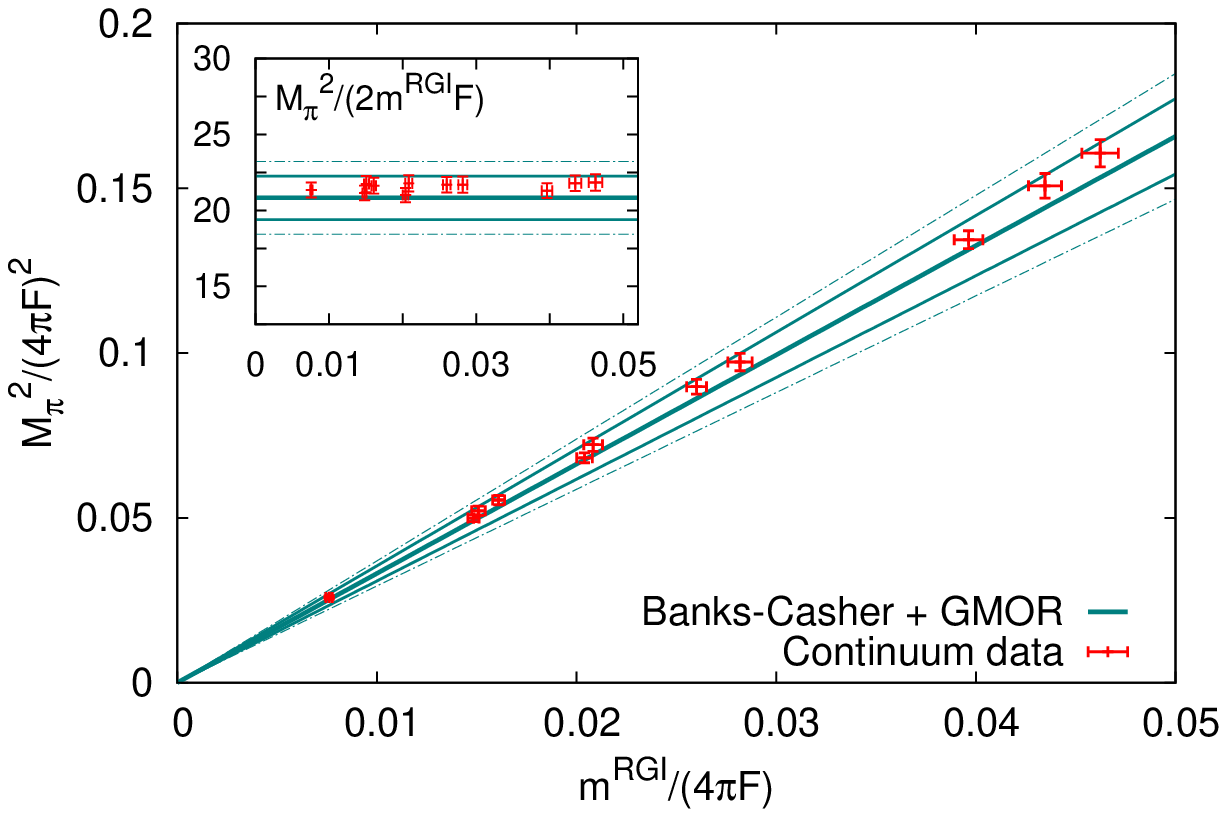}
\end{minipage}
\vspace{-0.25cm}

\caption{Left: effective spectral density $\wS$ in the continuum and chiral
limits. Right: the continuum extrapolated pion mass squared versus the RGI
quark mass, both normalized to $4 \pi F$. The central
line is the GMOR contribution to the pion mass squared computed by taking the 
direct measure of the condensate from the spectral density. 
The upper and lower solid lines show 
the statistical error and the dotted-dashed ones the total error, 
the systematic being added in quadrature. Courtesy of 
Refs.~\cite{Engel:2014cka,Engel:2014eea}.
}\label{fig:chiral_p_gmor}
\end{figure}

The extrapolation to the chiral limit requires an assumption on how the effective
spectral density behaves when $m \rightarrow 0$. The absence of chiral logarithms
in Eq.~(\ref{eq:RNLO}), however, implies that $\wS^{\rm nlo}$ is just a linear function of $m$ near
the origin. In this limit, as discussed in section~\ref{ssec:ChPT}, 
$\wS^{\rm nlo}=\Sigma$ holds also at non-zero $\Lambda$ since all
NLO corrections in Eq.~(\ref{eq:RNLO}) vanish. To check for this property, in
Refs.~\cite{Engel:2014cka,Engel:2014eea} the
data have been extrapolated following Eq.~(\ref{eq:RNLO}) but leaving the leading term
free to depend on $\Lambda$. The results of the fit are 
shown in the left plot of Figure~\ref{fig:chiral_p_gmor}. Within errors the
$\Lambda$-dependence is clearly compatible with a constant up to $\approx 80$~MeV.
Moreover the difference between the values of $\wS$ in the chiral limit and
those at $m=12.9$~MeV is of the order of the statistical error, i.e. the
extrapolation is very mild. A fit to a constant of the data gives~\cite{Engel:2014cka,Engel:2014eea}
\be\label{eq:bellarho}
[\pi\rho]^{1/3}=261(6)(8)~\mbox{MeV}\; , 
\ee
with the spacing being fixed in physical units by introducing a quenched
strange quark and requiring that $F_K=109.6$~MeV.\\

\begin{figure}[t!]
\hspace{-0.375cm}\begin{minipage}{0.35\textwidth}
\includegraphics[width=8.5 cm,angle=0]{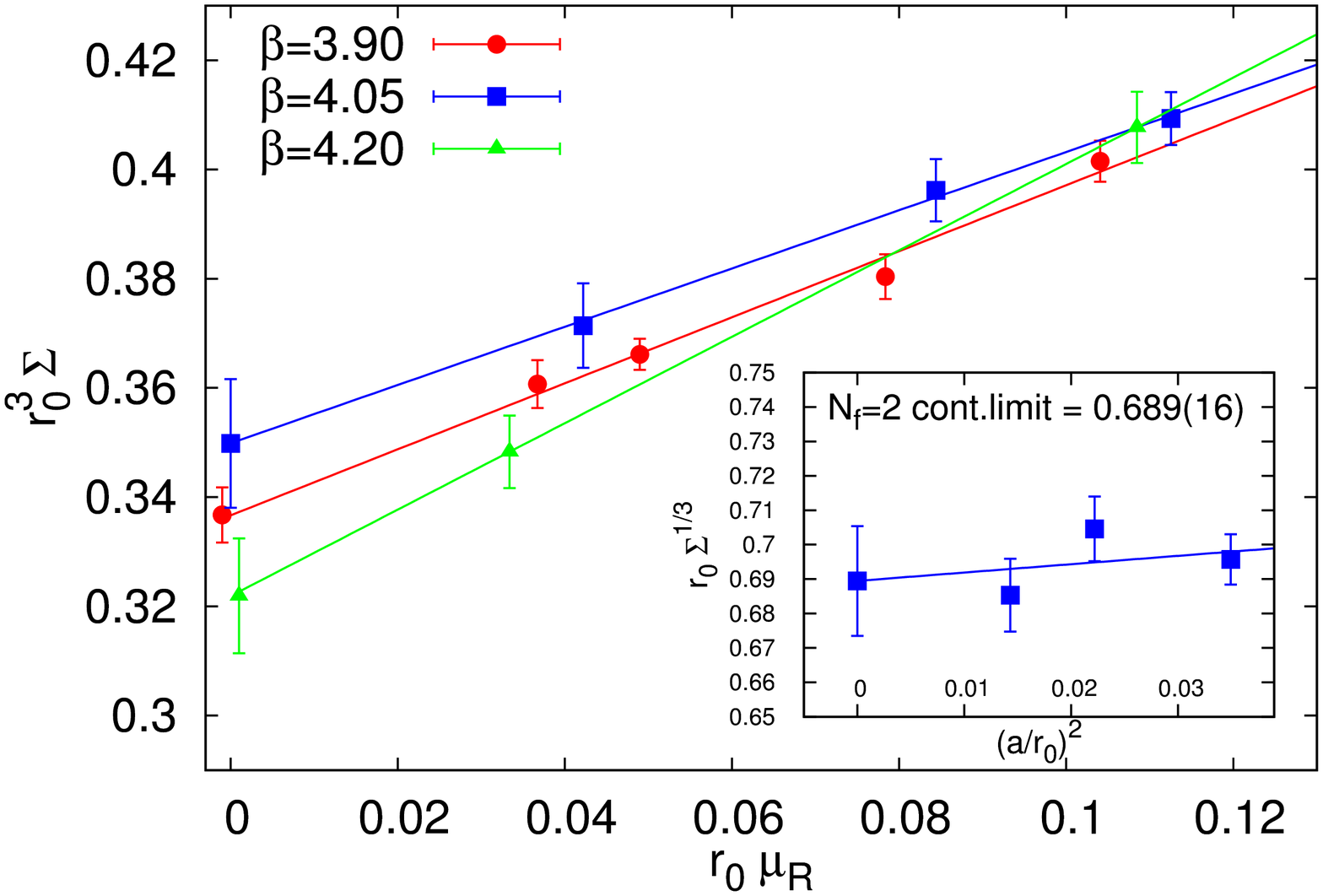}
\end{minipage}
\hspace{20mm}
\begin{minipage}{0.35\textwidth}
\includegraphics[width=8.5 cm,angle=0]{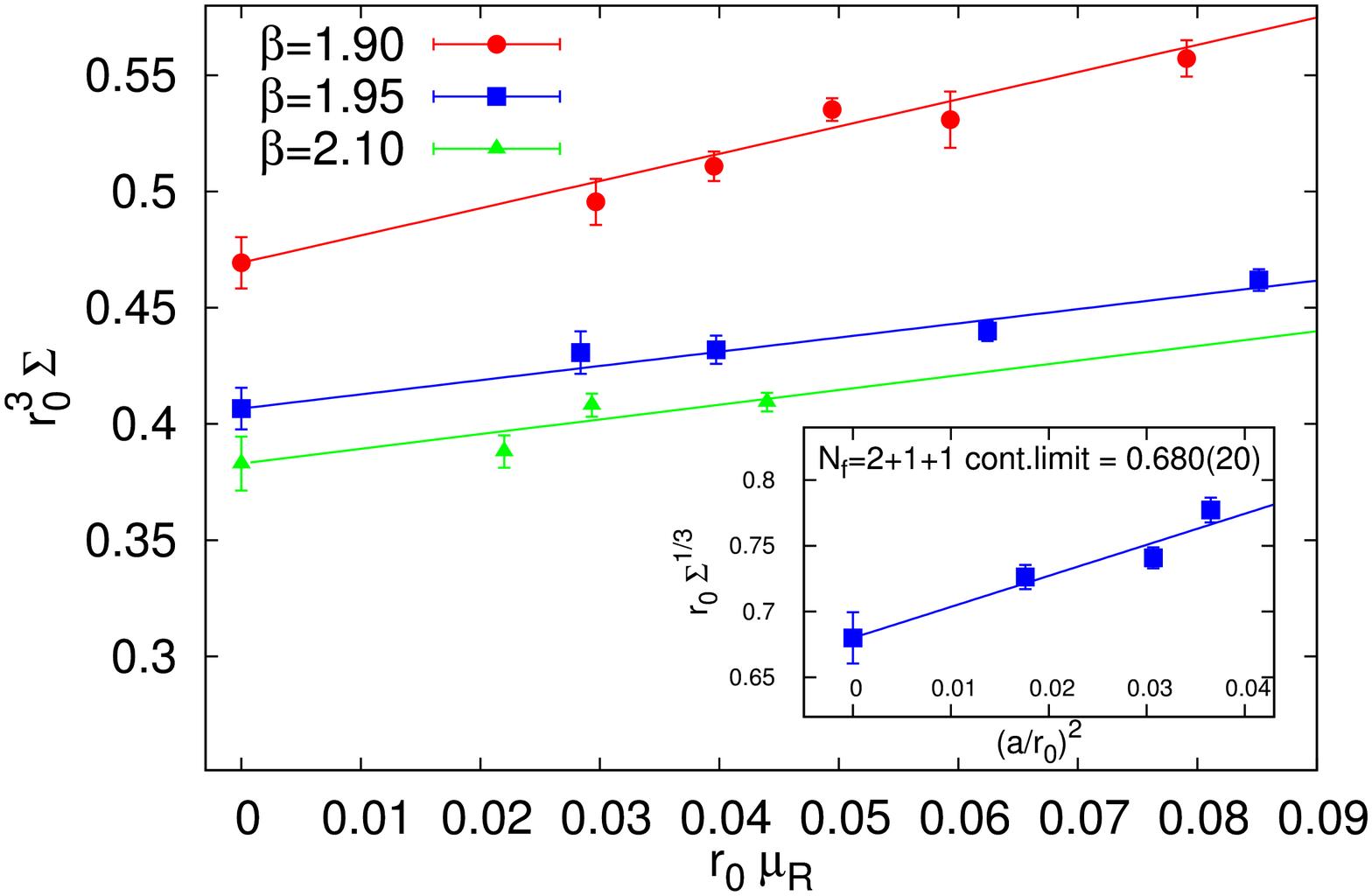}
\end{minipage}
\vspace{-0.75cm}

\caption{Chiral and continuum (inset) extrapolations of the spectral density for the $N_f=2$ (left)
and $N_f=2+1+1$ (right) ensembles from Ref.~\cite{Cichy:2013gja}. Courtesy of 
Ref.~\cite{Cichy:2013gja}.
\label{fig:fits_ETM}}
\end{figure}

The authors of Ref.~\cite{Cichy:2013gja} analyzed their data by following the 
strategy adopted in the exploratory computation of Ref.~\cite{Giusti:2008vb}. The mode
number at fixed lattice spacing and quark mass is fitted linearly in $M$ in the range
$50\leq M\leq 120$~MeV. The slope is then linearly extrapolated in $m$ to the chiral limit,
and the results are finally extrapolated to the continuum limit linearly in $a^2$,
see left plot in Figure~\ref{fig:fits_ETM}. The dependence of the slope on the fitting range is
estimated and included in the systematic error. The result that they get is
$r_0 [\pi\rho]^{1/3}=0.689(16)(29)$.
They prefer not to quote a value in MeV due to the large uncertainty in the
determination of the lattice spacing in physical units from the ETM Collaboration~\cite{Cichy:2013gja}. However,
if I use $r_0 F_K=0.2794(44)$ from \cite{Fritzsch:2012wq}, I get $[\pi\rho]^{1/3}=270(6)(11)(4)$~MeV,
an exercise that shows that their value is not inconsistent with the result in
Eq.~(\ref{eq:bellarho}). More work by the ETM Collaboration is desirable
to have a precise and reliable determination of the overall scale.
For completeness in the right plot of Figure~\ref{fig:fits_ETM}
the results for the analogous computation with $N_f=2+1+1$ is also
shown~\cite{Cichy:2013gja}. 

\subsection{GMOR relation}
As in any numerical computation, the chiral limit inevitably requires an 
extrapolation of the results with a pre-defined functional form. The distinctive
feature of spontaneous symmetry breaking, however, is that in the infinite volume
limit the value of $\pi \rho$ at the origin has to agree with the one of $M_\pi^2 F_\pi^2/2m$
in the chiral limit
\be
\lim_{m \to 0} \frac{M_\pi^2 F_\pi^2}{2m} = \pi \lim_{\lambda \to 0}\lim_{m \to 0}\rho(\lambda,m)\; . 
\ee
To make this comparison, the authors of Refs.~\cite{Engel:2014cka,Engel:2014eea}
complemented the computation of the mode number with those for the mass and the decay constant
of the pion, $M_\pi$ and $F_\pi$, as well as of the quark mass $m$. The dimensionless ratio
$M_\pi^2/2 m F$, extrapolated to the continuum limit, is shown in the right
plot of Figure~\ref{fig:chiral_p_gmor} together with the GMOR contribution to it
computed by taking the value of the spectral density in Eq.~(\ref{eq:bellarho}) (central line).
The upper and lower solid lines show the statistical
error and the dotted-dashed ones the total error, the systematic being added in quadrature.
A different way to appreciate the excellent agreement is to compare the value of
$[\pi\rho]^{1/3}$ reported in Eq.~(\ref{eq:bellarho}) with the one of the condensate extracted from
the GMOR relation which is $[\Sigma_{\rm GMOR}]^{1/3}= 263(3)(4)$~MeV. For completeness the value
obtained in Refs.~\cite{Engel:2014cka,Engel:2014eea} for the decay constant in the chiral limit
is $F=85.8(7)(20)$~MeV.

These results show that the spectral density of the Dirac operator in the continuum and chiral
limits is {\it non-zero}
at the origin. They provide a numerical proof of the fact that the low-modes of the Dirac
operator do condense as dictated by the Banks--Casher mechanism in presence of a non-zero chiral
condensate, and that the picture of spontaneously broken chiral symmetry in QCD is indeed correct.

A similar exercise can be done for the results in Ref.~\cite{Cichy:2013gja} and for those
of the ETM collaboration in Ref.~\cite{Baron:2009wt}. By taking $r_0=0.420(14)$ from~\cite{Baron:2009wt}
I obtain $[\pi\rho]^{1/3}=324(8)(14)(11)$, a value which has to be compared with
the condensate extracted from the overall fit of the pion mass and the decay constant
in the same reference, $\Sigma^{1/3}=269.9(65)$~MeV~\cite{Baron:2009wt}. The tension between the two values
is again not inconsistent with the conclusions drawn above due to the large uncertainty (larger
than the error quoted above) in the determination of $r_0$ in physical units
from the ETM Collaboration, and also due to the discrepancy with respect to the value given
before in units of $F_K$, see Refs.~\cite{Cichy:2013gja,Sommer:2014mea} for a detailed discussion. 

\subsection{Miscellaneous remarks}
From the analysis in Refs.~\cite{Engel:2014cka,Engel:2014eea}, the best results
for the leading-order low-energy constants of QCD with two flavours 
are
\be\label{eq:finalMeV}
    [\Sigma^\msbar(2\, \mbox{GeV})]^{1/3} =  263(3)(4)\; {\rm MeV}\; ,
    \quad F=85.8(7)(20)\; {\rm MeV}\; .
\ee
By taking the value of the $\Lambda$-parameter from the Alpha
collaboration~\cite{DellaMorte:2004bc,Fritzsch:2012wq}, and by
taking into account the correlation with $F$, the results above correspond to the dimensionless 
ratios~\cite{Engel:2014cka,Engel:2014eea}
\be\label{eq:bella}
\frac{[\Sigma^{\rm RGI}]^{1/3}}{F} =  2.77(2)(4)\;,\quad
\frac{\Lambda^{\msbar}}{F}  =  3.6(2)\; . 
\ee
where the renormalization-group-invariant (RGI) condensate 
is defined with the convention of Refs.~\cite{Capitani:1998mq,DellaMorte:2005kg}.
For the sake of clarity, often (also in the previous sections) the values of dimensionful quantities 
in the $N_f=2$ theory are expressed in physical units. They 
are, however, affected by an intrinsic ambiguity due to the matching of 
the quantity chosen to fix the lattice spacing with its experimental value. The 
renormalization group-invariant dimensionless ratios quoted in 
Eq.~(\ref{eq:bella}), however, 
are parameter-free predictions of the $N_f=2$ theory. They belong to the family 
of unambiguous quantities that should be used for comparing 
results in the two flavour theory at variance with what is usually
done in many reports, e.g.~\cite{Aoki:2013ldr}. 

The quantities in Eq.~(\ref{eq:bella}) can be used to confront predictions from models, 
large $N_c$ approximation, etc. with lattice results. In this respect it is 
interesting to note that several years ago an analytic computation of the chiral 
condensate in QCD was carried out in the context of the planar equivalence between 
the large $N_c$ limits of the ${\cal N}=1$ super Yang-Mills theory and a variant of QCD with 
fermions in the antisymmetric 
representation~\cite{Armoni:2003yv,Armoni:2005wt,Armoni:2014ywa}. In particular, an 
approximate expression for the quark condensate for $N_c=3$ QCD with quarks in the 
fundamental representation was inferred from an exact calculation of the gluino 
condensate in the ${\cal N}=1$ super Yang-Mills theory and further assumptions.
The results that these authors quote, with the renormalization 
conventions adopted in Eq.~(\ref{eq:bella}), is 
\be\label{eq:assv}
\frac{[\Sigma^{\rm RGI}]^{1/3}}{\Lambda^{\msbar}} = 1.43 
\Big[\frac{N_c}{2\pi^2} K_{\rm F}(1/N_c,N_f)\Big]^{1/3}
\ee
where the numerical pre-factor $1.43$ is just conventional and takes into account of the 
different conventions adopted in Refs.~\cite{Armoni:2003yv,Armoni:2005wt,Armoni:2014ywa} with
respect to those in Eq.~(\ref{eq:bella}). The $K_{\rm F}$ function encodes the sub-dominant 
$1/N_c$ corrections, which they conjecture to be relatively small. If I now compare the expression 
in Eq.~(\ref{eq:assv}) with the 
results in Eq.~(\ref{eq:bella}), I obtain $[K_{\rm F}(1/3,2)\Big]^{1/3}=1.01(5)$ which, if taken at
face value, is indeed compatible with having small sub-dominant corrections.
Further lattice computations at various $N_c$ and $N_f$ are needed to test the conjecture,
but the comparison above provides a good example on how the results in Eq.~(\ref{eq:bella}) can be used to
test predictions from models and/or analytic approximations.

\section{The Witten--Veneziano relation and the gradient-flow} %
The topological susceptibility in the SU($N_c$) Yang--Mills theory
can be formally defined in Euclidean space-time as
\begin{equation}
\chi = \int d^4x\: 
\langle q(x) q(0) \rangle \; ,
\label{eq:chidef}
\end{equation}
where the topological charge density $q(x)$ is
\begin{equation}\label{eq:naive}
q(x)=\frac{1}{64\pi^2} 
\epsilon_{\mu\nu\rho\sigma} F^a_{\mu\nu}(x) F^a_{\rho\sigma}(x)\; .
\end{equation}
In general the definition (\ref{eq:naive}) of the topological charge density 
needs to be combined with an unambiguous renormalization condition. The cumulants of
the charge, e.g. the susceptibility in Eq.~(\ref{eq:chidef}), require also additional
subtractions of short-distance singularities to make the correlators of the
charge density integrable distributions.
Besides its interest within the pure gauge theory, $\chi$
plays a crucial r\^ole in the QCD-based explanation of the large 
mass of the $\eta'$ meson proposed  by Witten and Veneziano (WV) 
\cite{Witten:1979vv,Veneziano:1979ec}. Analogously to the
GMOR relation, the WV mechanism
predicts that at leading order in $1/N_\mathrm{c}$
the contribution due to the anomaly to the mass of the pseudoscalar meson
associated to the generator of $U(1)_A$ is 
given by \cite{Witten:1979vv,Veneziano:1979ec,Giusti:2001xh,Seiler:2001je} 
\begin{equation}
\displaystyle
\lim_{m \to 0}\; \lim_{N_c\to \infty} 
\frac{F_{\pi}^2 m^2_{\eta^\prime}}{2 N_\mathrm{f}} = \lim_{N_c\to \infty}  \chi\, .  
\label{eq:WV}
\end{equation}
For the Eq.~(\ref{eq:WV}) to be valid, the renormalization
conditions for the charge and for the susceptibility need to be chosen so that in presence of fermions
the anomalous chiral Ward identities are satisfied~\cite{Giusti:2001xh,Giusti:2004qd,Luscher:2004fu}.
In this case the value of $\chi$ when $N_c\to \infty$ corresponds also to
a low-energy constant in the simultaneous expansion in momenta and in $1/N_c$ of
the U$(3)$ chiral effective theory~\cite{DiVecchia:1980ve,Witten:1980sp,DiVecchia:1980sq,Kaiser:2000gs}.

We know three families of definitions of the topological charge whose
cumulants are ultraviolet finite and unambiguous: the one suggested
by Ginsparg--Wilson (GW) fermions
\cite{Neuberger:1997fp,Hasenfratz:1998ri,Luscher:1998pqa,Giusti:2001xh,Giusti:2004qd,Luscher:2004fu},
the spectral projector formulas~\cite{Giusti:2008vb}, and the naive definition at
positive flow time~\cite{Luscher:2010iy}. When the topological charge is defined as suggested
by GW, its bare lattice expression and those of the corresponding cumulants have finite and
unambiguous continuum limits as they stand which satisfy the anomalous 
chiral Ward identities by construction. The continuum limit of this  definition of $\chi$
is the right one to be inserted in Eq.~(\ref{eq:WV}).
By combining a series of anomalous chiral Ward identities, the cumulants can
be written as integrated correlation functions of scalar and pseudoscalalar density
chains~\cite{Giusti:2004qd,Luscher:2004fu}, and
well chosen combinations of them correspond to the spectral projector definitions in
Ref.~\cite{Giusti:2008vb}. Written in this form, a GW regularization is not required anymore to
prove that no renormalization factor or subtractions of short-distance singularities are required.
The spectral projector expressions thus provide a universal definition of the susceptibility and
of the higher cumulants
which, in the continuum limit, satisfy the anomalous chiral Ward Identities since in a chirally symmetric
regularization they coincide with the GW ones.

Recently a third family of definitions of the topological charge was 
found~\cite{Luscher:2010iy}, whose cumulants have a finite and unambiguous
continuum limit~\cite{Luscher:2010iy,Luscher:2011bx}. It is a naive 
discretization of the charge evolved with the Yang--Mills gradient-flow. 
It is particularly appealing because its numerical evaluation is significantly
cheaper than the others.
This year there has been progress in this direction, and
it has been shown that indeed the continuum limit of the cumulants of topological
charge defined by the gradient-flow at positive flow time coincide with
those of the universal definition. After more than 30 years, the long story of defining the
topological charge on the lattice comes to a satisfactory end with a simple and elegant solution.

\subsection{Gradient-flow definition of the topological charge in the continuum\label{ssec:cont}}
Starting from the ordinary fundamental gauge field 
\be
B_\mu \Big|_{t=0} = A_\mu\; , 
\ee
the Yang--Mills gradient-flow evolves the gauge field as a function
of the flow time $t\geq 0$ by solving the differential
equation~\cite{Luscher:2010iy} 
\bea
\partial_t B_\mu & = & D_\nu G_{\nu\mu} + \alpha_0 D_\mu \partial_\nu B_\nu\; , \\[0.25cm]
G_{\mu\nu} & = & \partial_\mu B_\nu - \partial_\nu B_\mu - i[B_\mu,B_\nu]\; , \qquad
D_\mu = \partial_\mu - i\, [B_\mu,\cdot]\; ,
\eea
with $\alpha_0$ being the parameter which determines the gauge. Here we focus on 
the gradient-flow evolution of the topological charge density defined 
as
\be\label{qtcont}
q^t = \frac{1}{64 \pi^2} \epsilon_{\mu\nu\rho\sigma} G^a_{\mu\nu} G^a_{\rho\sigma} \; , 
\ee
and of the corresponding topological charge
\be
Q^t = \int d^4 x\, q^t(x)\; .  
\ee
When $q^t(x)$ is inserted in a correlation function
at a physical distance with any finite (multi)local
operator $O(y)$, it holds~\cite{Ce:2015qha}
\be\label{eq:lhsexp}
\langle q^t(x)\, O(y) \rangle = \langle q^{t=0}(x)\, O(y) \rangle + 
\partial_\rho \int_0^t dt'\, \langle w^{t'}_\rho(x)\, O(y)\rangle
\qquad (x \neq y;\; \rho=0,\dots,3)\; ,
\ee
where $w^t_\rho$ is a dimension-$5$ gauge-invariant pseudovector field.
The l.h.s. of Eq.~(\ref{eq:lhsexp}) is finite 
thanks to the fact that a gauge-invariant local composite 
field constructed with the gauge field evolved at positive flow time 
is finite~\cite{Luscher:2010iy,Luscher:2011bx}. Since there are no local composite
fields of dimension $d<5$ with the symmetry properties of 
$w^t_{\rho}(x)$, the integrand on the r.h.s 
of Eq.~(\ref{eq:lhsexp}) diverges at most logarithmically when $t'\rightarrow 0$. This 
implies that the quantity
\be\label{eq:qt0def}
\langle q^{t=0}(x)\, O(y) \rangle \equiv \lim_{t\rightarrow 0}\, \langle q^t(x)\, O(y) \rangle \qquad (x \neq y)\; ,   
\ee
is finite, i.e. the limit on the r.h.s exists for any finite operator $O(y)$. The 
Eq.~(\ref{eq:qt0def}) can be taken as the definition of $q^{t=0}(x)$, i.e. the renormalized 
topological charge density operator at $t=0$.
It is worth noting that Eq.~(\ref{eq:lhsexp}) implies that the small-$t$ 
expansion of $q^t(x)$ is of the form 
\be\label{eq:smallt}
\langle q^t(x)\, O(y) \rangle = 
\langle q^{t=0}(x)\, O(y) \rangle + {\cal O}(t) \qquad (x \neq y)\; ,  
\ee
with no divergences when $t\rightarrow 0$.

In the following we are interested in supplementing the theory with extra degenerate valence quarks
of mass $m$, and in considering the (integrated) correlator of a topological charge
density with a chain made of scalar and pseudoscalar densities
defined as~\cite{Luscher:2004fu}  
\be\label{eq:chain_cont}
\hspace{-0.325cm}  
\langle q^{t=0}(0) P_{51}(z_1) S_{12}(z_2) S_{23}(z_3) S_{34}(z_4) S_{45}(z_5) \rangle\; , 
\ee
where $S_{ij}$ and $P_{ij}$ are the scalar and the pseudoscalar renormalized densities with
flavor indices $i$ 
and $j$. Power counting and the operator product expansion predict that there are no
non-integrable short-distance singularities when the coordinates of two or
more densities in (\ref{eq:chain_cont}) tend to coincide among themselves
or with $0$. When only one of the densities is close to $q^{t=0}(0) $,
the operator product expansion predicts the leading singularity to be  
\be\label{eq:leadingsds}
q^{t=0}(x)\, S_{ij}(0) \xrightarrow{x\to\, 0}
c(x)\, P_{ij}(0) + \dots
\ee
where $c(x)$ is a function which diverges as 
$|x|^{-4}$ when $|x|\rightarrow 0$, and the dots indicate 
sub-leading contributions. An analogous expression 
is valid for the pseudoscalar density. Being the leading short-distance singularity 
in the product of fields $q^{t=0}(x)\, S_{ij}(0)$, 
its Wilson coefficient $c(x)$  can be computed in perturbation theory. By using 
Eq.~(\ref{eq:lhsexp}), to all orders in perturbation
theory
we can write
\be\label{eq:bellissima}
\langle q^{t=0}(x) S_{ij}(0)\, O(y) \rangle = 
\langle q^{t}(x) S_{ij}(0)\, O(y) \rangle - 
\partial_\rho \int_0^t dt' 
\langle w^{t'}_{\rho}(x) S_{ij}(0) \, O(y) \rangle \; ,
\ee
where again $O(y)$ is any finite (multi)local operator inserted at a physical 
distance from $0$ and $x$. When $t>0$, the first member on the r.h.s 
of Eq.~(\ref{eq:bellissima}) has no singularities when $|x|\rightarrow 0$.
If present, the singularity has to come from the second term, and therefore
$c(x)$ must be of the form 
\be\label{eq:bellissima2}
c(x) = \partial_\rho u_\rho(x)
\ee
which does not contribute to the integral (over all coordinates) of the 
correlation function (\ref{eq:chain_cont}).

\subsection{Ginsparg--Wilson definition of the charge density with the gradient-flow}
The definition of the topological charge density suggested 
by Ginsparg--Wilson fermions is~\cite{Neuberger:1997fp,Luscher:1998pqa,Hasenfratz:1998ri}
\begin{equation}\label{eq:qx}
a^4 q^t_\textsc{n}(x) = -\frac{\bar a}{2}\, \tr\Big[\gamma_5 D(x,x)\Big] , 
\end{equation}
where we indicate it with a subscript $\textsc{n}$ since, for concreteness,
we take $D(x,y)$ to be the Neuberger--Dirac operator in which each link variable 
is replaced by the corresponding evolved one when $t>0$. 
Since there are no other operators of dimension $d\leq 4$ which are 
pseudoscalar and gauge-invariant, it holds that 
\be\label{eq:zq1}
\lim_{a\rightarrow 0} Z_q\, \langle q^t_\textsc{n}(0)\, q^{t=0}_\textsc{n}(x) \rangle = 
{\rm finite}\; ,  
\ee
where $Z_q$ is a renormalization constant which is at most logarithmically divergent,
while $q^t_\textsc{n}(0)$ is finite as it stands. This in turn implies that 
\be\label{eq:cd1}
\lim_{a\rightarrow 0} Z_q\, a^4 \sum_x\, \langle q^t_\textsc{n}(0)\, q^{t=0}_\textsc{n}(x) 
\rangle = {\rm finite}\; ,
\ee
since there are no short-distance singularities that contribute to the integrated 
correlation function because $q^t_\textsc{n}(0)$ is evolved at positive flow-time. 
By supplementing the theory  with extra 
degenerate valence quarks of mass $m$, and by replacing in Eq.~(\ref{eq:cd1}) 
the topological charge at $t=0$ with its density-chain 
expression we obtain
\be\label{eq:chain1}
a^4 \sum_x \langle q^{t}_\textsc{n}(0)\, q^{t=0}_\textsc{n}(x)\rangle  = -m^5 a^{20}
\sum_{z_1,\dots,z_5}
\langle q^{t}_\textsc{n}(0)\, P_{51}(z_1)\, S_{12}(z_2)\, S_{23}(z_3)\, S_{34}(z_4)\, S_{45}(z_5) 
\rangle\; . 
\ee 
Written as in Eq.~(\ref{eq:chain1}), power counting and the operator product expansion predict 
that there are no non-integrable short-distance singularities when the coordinates 
of two or more densities tend to coincide. The r.h.s of Eq.~(\ref{eq:chain1}) is finite as 
it stands, and it converges to the continuum limit with a rate proportional to $a^2$.
This in turn implies that the limits
on the l.h.s of Eqs.~(\ref{eq:zq1}) and (\ref{eq:cd1}) are reached with the same rate
if $Z_q$ is set to any fixed ($g_0$-independent) value. Since in the classical continuum
limit Neuberger's definition 
in Eq.~(\ref{eq:qx}) tends to the one in 
Eq.~(\ref{qtcont}) \cite{Kikukawa:1998pd,Fujikawa:1998if}, we may set
$Z_q=1$ in which case
\be\label{eq:bellaq}
\lim_{a\rightarrow 0}\, \langle q^{t=0}_\textsc{n}(x)\, O_\textsc{l}(y) \rangle  =
\langle q^{t=0}(x)\, O(y) \rangle \qquad (x \neq y)\; ,
\ee
where $O_\textsc{l}(y)$ is a discretization of the generic finite continuum operator $O(y)$. 
Once inserted in correlation functions at a physical distance from 
other (renormalized) fields, $q^{t=0}_\textsc{n}(x)$ does not require any renormalization 
in the Yang--Mills theory. It is finite as it stands, and it satisfies the singlet Ward 
identities when fermions are included in the theory. So does $q^{t=0}(x)$.

\subsection{Ginsparg--Wilson definition of the charge cumulants}
The Neuberger's definition of the topological charge is given by 
\be
Q^t_\textsc{n} \equiv a^4 \sum_x q^t_\textsc{n}(x)\; , 
\ee  
and its cumulants are defined as 
\be\label{eq:CnQCd}
{C}^t_{\textsc{n},n} = a^{8n-4}\!\!\!\!\!\!\sum_{x_1,\dots,x_{2n-1}}
\langle q^t_\textsc{n}(x_1)\dots q^t_\textsc{n}(x_{2n-1})\, q^t_\textsc{n}(0)\rangle_\mathrm{c} \; .
\ee
For $t=0$ the cumulants have an unambiguous universal continuum limit as they stand and,
when fermions are included, they satisfy the proper singlet chiral 
Ward identities~\cite{Giusti:2001xh,Giusti:2004qd,Luscher:2004fu}.
It is far from being obvious that ${C}^{t=0}_{\textsc{n},n}$ coincide with those defined at 
positive flow-time, since the two definitions may differ by additional finite 
contributions from short-distance singularities.

For the clarity of the presentation we start by focusing on the lowest cumulant, 
the topological susceptibility ${C}^{t}_{\textsc{n},1}$. At $t=0$, by replacing one 
of the two $q^{t=0}_\textsc{n}$ 
with its density-chain expression~\cite{Luscher:2004fu}, we obtain
\be\label{eq:chain}
a^4 \sum_x \langle q^{t=0}_\textsc{n}(0) q^{t=0}_\textsc{n}(x)\rangle  = -m^5 a^{20}
\sum_{z_1,\dots,z_5}
\langle q^{t=0}_\textsc{n}(0) P_{51}(z_1) S_{12}(z_2) S_{23}(z_3) S_{34}(z_4) S_{45}(z_5) \rangle\; . 
\ee
When the susceptibility is written in this form, the discussion toward the end of
section \ref{ssec:cont} and in particular Eq.~(\ref{eq:bellissima2}) guarantee that
there are no contributions from short-distance singularities. This result, together with
the fact that $Z_q=1$, implies that  
\be\label{eq:chit0}
\lim_{t\rightarrow 0}\, \lim_{a\rightarrow 0} a^4 \sum_x \langle q^{t}_\textsc{n}(x)\, q^{t=0}_\textsc{n}(0)\rangle = 
\lim_{a\rightarrow 0} a^4 \sum_x \langle q^{t=0}_\textsc{n}(x)\, q^{t=0}_\textsc{n}(0)\rangle \; .  
\ee
By replacing on the l.h.s $q^{t=0}_\textsc{n}(0)$ with the evolved one, no further 
short-distance singularities are introduced and we arrive to the final result
\be\label{eq:finalres}
\lim_{t\rightarrow 0}\, \lim_{a\rightarrow 0} a^4 \sum_x 
\langle q^{t}_\textsc{n}(x)\, q^{t}_\textsc{n}(0)\rangle = 
\lim_{a\rightarrow 0} a^4 \sum_x \langle q^{t=0}_\textsc{n}(x)\, q^{t=0}_\textsc{n}(0)\rangle \; .
\ee

By replacing $2n-1$ of the charges in the $n^{th}$ cumulant with their density-chain 
definitions, the very same line of argumentation can be applied. The 
Eq.~(\ref{eq:finalres}), together with the independence up to harmless discretization effects 
of ${C}^t_{\textsc{n},n}$ from the flow-time for $t>0$~\cite{Luscher:2010iy},
implies that the continuum limit of 
${C}^t_{\textsc{n},n}$ coincides with the one of ${C}^{t=0}_{\textsc{n},n}$. The cumulants
of the topological charge distribution defined at $t>0$ thus 
satisfy the proper singlet chiral Ward identities when fermions are 
included.
They are the proper quantities to be inserted in the Witten--Veneziano relations 
for the mass and scattering amplitudes of the $\eta'$ meson in 
QCD.

\subsection{Universality at positive flow-time \label{sec:naive}}
For $t>0$ different lattice definitions of the topological charge density 
tend to the same continuum limit if they share the same asymptotic 
behavior in the classical continuum limit~\cite{Luscher:2010iy,Luscher:2011bx}.
We can therefore consider the naive definition of the topological charge
density defined as
\begin{equation}
\label{eq:topological_charge}
q^t(x) = \frac{1}{64\pi^2}\, 
\epsilon_{\mu\nu\rho\sigma}\, G_{\mu\nu}^a(x) G_{\rho\sigma}^a(x)\; ,
\end{equation}
where the field strength tensor $G^a_{\mu\nu}(x)$ is defined 
as
\begin{equation}
\label{eq:field_strength_tensor_clover}
G_{\mu\nu}^a(x) = - \frac{i}{4 a^2}
\tr[\left( Q_{\mu\nu}(x) - Q_{\nu\mu}(x) \right)\, T^a ]\; , 
\end{equation}
with 
\bea
    Q_{\mu\nu}(x) & = &\, V_\mu(x) V_\nu(x + a\dmu) \Vd_\mu(x + a\dnu) \Vd_\nu(x) + \nonumber\\[0.125cm]
               & &\, V_\nu(x) \Vd_\mu(x - a\dmu + a\dnu) \Vd_\nu(x - a\dmu) V_\mu(x - a\dmu) +\nonumber\\[0.125cm]
 & &\, \Vd_\mu(x - a\dmu) \Vd_\nu(x - a\dmu - a\dnu) V_\mu(x - a\dmu - a\dnu) V_\nu(x - a\dnu) + \\[0.125cm]
 & &\, \Vd_\nu(x - a\dnu) V_\mu(x - a\dnu) V_\nu(x + a\dmu - a\dnu) \Vd_\mu(x)\; .\nonumber     
\eea
In the Yang--Mills theory $q^{t=0}(x)$ requires a multiplicative renormalization
constant when inserted in correlation 
functions at a physical distance from other 
operators \cite{Alles:1996nm}. The cumulants of the corresponding topological 
charge, defined analogously to 
Eq.~(\ref{eq:CnQCd}), have additional ultraviolet power-divergent singularities, 
and they do not have a continuum limit. 

The density $q^{t}(x)$ in 
Eq.~(\ref{eq:topological_charge}) shares with $q^{t}_\textsc{n}(x)$  
the same asymptotic behavior in the classical 
continuum limit~\cite{Kikukawa:1998pd,Fujikawa:1998if}.
Since for $t>0$ short-distance singularities cannot arise, 
$C^t_{\textsc{n},n}$ and $C^t_{n}$ tend to the same continuum limit. 
The results in the previous section then imply that the continuum limit of the 
naive definition of $C^t_n$, {\it at positive flow-time}, coincides with the 
universal definition which satisfies the chiral Ward identities when fermions 
are added~\cite{Giusti:2001xh,Giusti:2004qd,Luscher:2004fu}. It is interesting to note, however, that 
at fixed lattice spacing there can be quite some differences. For instance, the topological 
susceptibility defined at $t>0$ with the naive definition is not guaranteed to go to zero in the 
chiral limit at finite lattice spacing in presence of fermions~\cite{Bruno:2014ova}. 

\section{First two cumulants of the charge distribution in the SU(3) Yang--Mills theory}
\begin{figure}[t!]
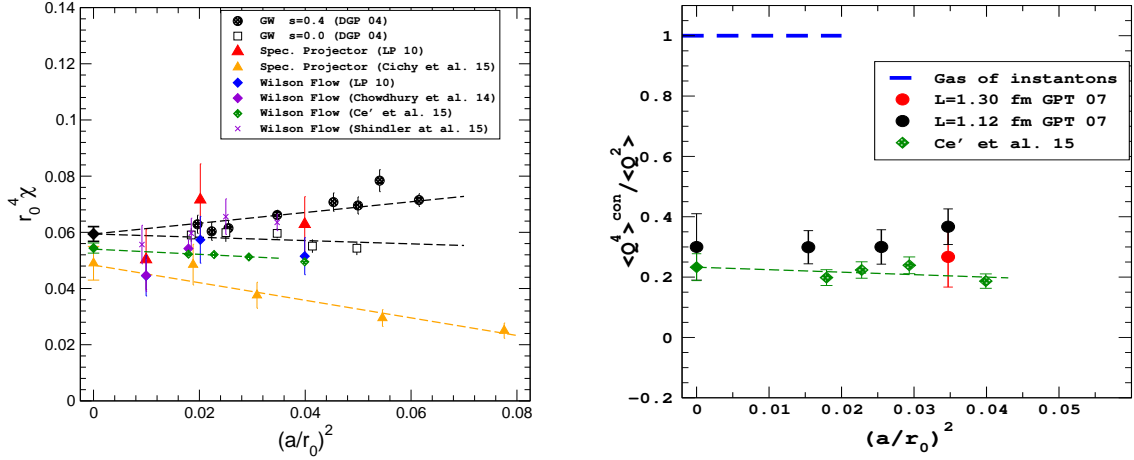

\vspace{0.25cm}
  
\begin{minipage}{0.35\textwidth}
  
\includegraphics[width=7.0 cm,angle=0]{./Q2_plot_universality.eps}
\end{minipage}
\hspace{25mm}
\begin{minipage}{0.35\textwidth}
  
\includegraphics[width=7.0 cm,height=6.0 cm,angle=0]{./Q4_r0a.eps}
\end{minipage}
\caption{The topological susceptibility $\chi$ (left) and the ratio $R=\langle Q^4\rangle_c/\langle Q^2\rangle$ in the SU($3$)
  Yang--Mills theory versus $(a/r_0)^2$.
}
\label{fig:cumssu3}
\end{figure}

Over the last year or so the topological susceptibility of the SU($3$)
Yang--Mills theory has been computed by several groups on a rich set of lattices with
a statistical error of a few percent
~\cite{Chowdhury:2013mea,Cichy:2015jra,Ce:2015qha,Shindler:2015aqa}. The results
are shown in the left plot of Figure~\ref{fig:cumssu3} together with the older results
obtained with the Neuberger~\cite{DelDebbio:2004ns} and
with the spectral projector~\cite{Luscher:2010ik} definitions of the charge. The best continuum
limit values from the three families of definitions are\footnote{Unless explicitly indicated, the gradient
flow-time at which the topological quantities are computed throughout this and the next section is at $t=t_0$.}
\bea
r_0^4 \chi & = & 0.0544\pm 0.0018\quad \mbox{gradient-flow~\cite{Ce:2015qha}}\\
r_0^4 \chi & = & 0.059\pm 0.003\quad \;\;\;\;\;\mbox{Neuberger definition~\cite{DelDebbio:2004ns}}\\
r_0^4 \chi & = & 0.049\pm 0.006\quad \;\;\;\;\;\mbox{modified spectral projector~\cite{Cichy:2015jra}}
\; , \nonumber
\eea
see left plot in Figure~\ref{fig:cumssu3}. They all agree within less than 2 standard deviations.
By computing the topological susceptibility at different flow times, one can
determine the ratios $\chi_t/\chi$ for which the statistical correlations among data
reduce the error significantly. In Ref.~\cite{Ce:2015qha} the continuum limit of
this ratio was found to be compatible with $1$ at the permille level at various values of $t$, another 
universality test far from being trivial. At finite lattice spacing, in fact, discretization effects are
clearly visible, and they depend on $t$. All these numerical results are consistent with the conceptual
progress made over the last decade. Up to date there is no sign of non-universal behaviour of
$\chi$ in the continuum limit, if the topological charge is properly defined on
the lattice. We have moved from an unsolved problem in quantum field theory to precise universality
tests! 

The numerical computation of the charge defined by the gradient-flow is orders of magnitude
cheaper than for the other two expressions. Since its cumulants at positive flow time coincide
with those of the universal definition and discretization effects remain mild, the flow
expression becomes the way to go for further (precise) studies of topological observables on the lattice.
Given the statistical precision that can be easily reached, 
the all-time favorite Sommer reference scale $r_0$~\cite{Sommer:1993ce} needs to be replaced by
quantities which can be determined with a higher statistical (and systematic) precision, such
as those derived from the gradient-flow~\cite{Sommer:2014mea}. By using $t_0$
as a reference scale~\cite{Luscher:2010iy}, the authors of Ref.~\cite{Ce:2015qha} obtain the most
precise determination of the
topological susceptibility in the SU($3$) Yang--Mills theory to date
\[
t_0^2 \chi = (6.67 \pm 0.07)\cdot 10^{-4}\; , 
\]
a value which is almost 3 times more precise than the one that can be obtained by using the
present best determination of $r_0$.

In Ref.~\cite{Ce:2015qha} the ratio of the second cumulant over the topological susceptibility,
$R=\langle Q^4\rangle_c/\langle Q^2\rangle$, was also computed with the gradient-flow definition
by keeping for the first time all systematics, especially finite volume effects, negligible with
respect to the statistical errors. The results are shown in green on the
right plot of Fig.~\ref{fig:cumssu3}. Their continuum limit extrapolation is
\be\label{eq:R}
R  = 0.233 \pm 0.045
\ee
which is in agreement with a previous determination obtained with the GW
definition~\cite{Giusti:2007tu}, black and red points in the right plot
of Fig.~\ref{fig:cumssu3}, albeit with an error 2.5 times smaller.
The result in Eq.~(\ref{eq:R}) is incompatible with the $\theta$-behaviour
of the vacuum energy predicted by dilute instanton models, for 
which $R=1$ \cite{Callan:1977gz}. It suggests that the quantum fluctuations of the 
topological charge are of quantum non-perturbative nature in the ensemble 
of gauge configurations that dominate the path integral.
The large $N_c$ expansion does not provide a sharp prediction for the value 
of $R$.  Its small value, however, is compatible with being a quantity suppressed
as $1/N_c^2$ in the large $N_c$ limit.

\section{Topological susceptibility in QCD}
The computation of the topological susceptibility is expensive
with respect to other quantities computed in (full) QCD. 
This is due to the long autocorrelation encountered in the
simulations~\cite{DelDebbio:2002xa,Schaefer:2010hu,Luscher:2011kk}, 
and to the intrinsic fluctuations of this quantity. By expanding the charge
distribution around the leading Gaussian behaviour, the Edgeworth expansion
leads to the estimate of the relative error for the susceptibility given by 
$\Delta \chi/\chi = \sqrt{2/N_{\rm conf}}+\dots$, where the dots indicate terms
which are suppressed by powers of $1/V$~\cite{Giusti:2007tu}. For a precision
of $\sim 5\%$, ${\cal O}(1000)$ independent configurations are needed.

In the presence of spontaneous symmetry breaking with $N_f$ degenerate light
flavours, the topological susceptibility toward the chiral limit goes as
\be
\chi_{\rm QCD} = \frac{\Sigma}{N_f} m + {\cal O}(m^2)\; , 
\ee
which implies a significant suppression with respect to the Yang--Mills theory.
Moreover being a pure gluonic operator,
the slope of $\chi_{\rm QCD}$ is a measurement a posteriori of the number of flavours simulated
once the chiral condensate is known. Vice versa, given the cost of its computation, 
the topological susceptibility is not a competitive quantity to determine $\Sigma$.

\begin{figure}[t!]
  
\begin{minipage}{0.35\textwidth}

\includegraphics[width=7.0 cm,angle=0]{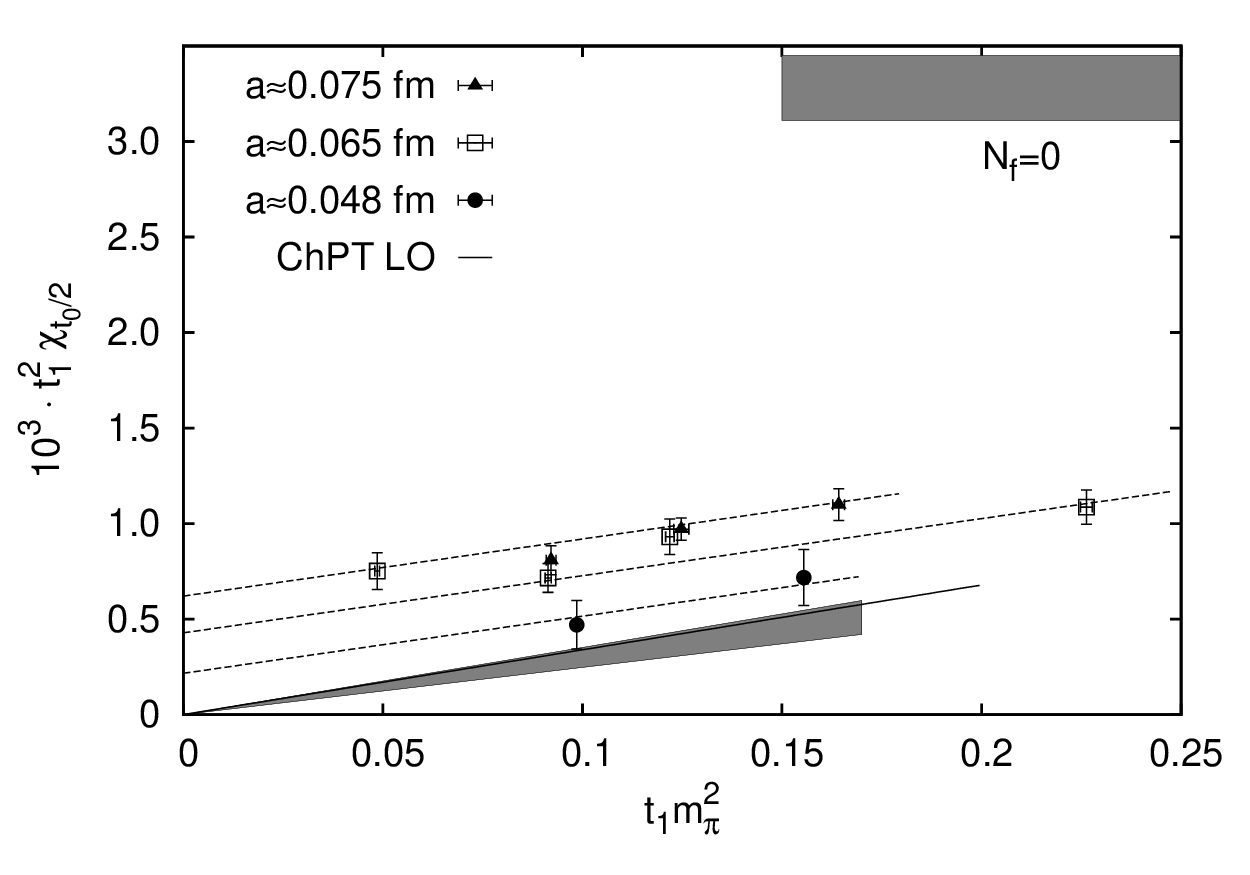}

\end{minipage}
\hspace{20mm}
\begin{minipage}{0.35\textwidth}

\includegraphics[width=7.0 cm,height=4.9 cm,angle=0]{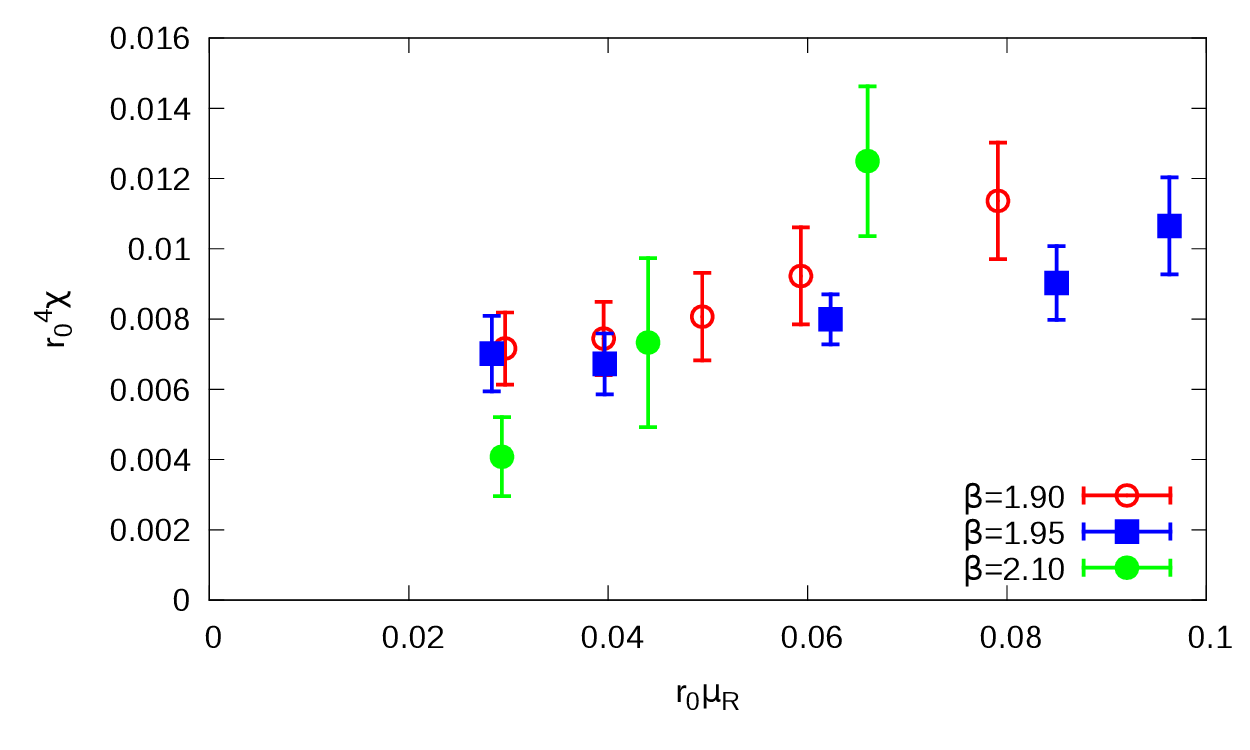}

\end{minipage}
\vspace{-0.375cm}

\caption{Left: topological susceptibility of the $N_f=2$ theory
  for all ensembles considered in Ref.~\cite{Bruno:2014ova}.
  Right: the susceptibility for the $N_f=2+1+1$ theory 
  for all ensembles simulated in Ref.~\cite{Cichy:2013rra}. 
Courtesy of Refs.~\cite{Bruno:2014ova,Cichy:2013rra}.}
\label{fig:chiQCDfull}
\end{figure}

In the last couple of years there have been two computations of $\chi_{\rm QCD}$, both
carried out on lattices with ${\cal O}(100)$ independent gauge
configurations\footnote{A recent attempt to extract the topological
susceptibility from the local fluctuations of the charge density two-point correlator
can be found in Ref.~\cite{Fukaya:2014zda}.}.
The statistical errors are therefore still quite large. 
The Alpha collaboration computed the susceptibility, defined from the gradient-flow
charge density as in Eqs.~(\ref{eq:topological_charge}), in the $N_f=2$ theory discretized 
with the standard Wilson gluonic action and the non-perturbatively
${\cal O}(a)$-improved Wilson fermion action. They used the gauge configurations
generated by the CLS community and by themselves to obtain
the results shown in the left plot of Figure~\ref{fig:chiQCDfull}~\cite{Bruno:2014ova}.
The ETM collaboration computed the susceptibility by implementing a variant 
of the spectral projection definition in the $N_f=2$ and 
$N_f=2+1+1$ theories. They have simulated the tree-level Symanzink ($N_f=2$) and
the Iwasaki gluon actions ($N_f=2+1+1$), and the Wilson twisted mass
fermion action. The results that they have obtained in the
$N_f=2+1+1$ theory is shown in the right plot of Figure~\ref{fig:chiQCDfull} as a
function of the renormalized quark mass~\cite{Cichy:2013rra}.

The suppression of $\chi_{\rm QCD}$ with respect to the Yang--Mills result is
clearly seen in both sets of results, and it is manifest in the left plot where also
the Yang--Mills value of the susceptibility is reported. In either cases there is no
reason for $\chi_{\rm QCD}$ to vanish in the chiral limit at finite lattice spacing. The authors
of Ref.~\cite{Bruno:2014ova} indeed extrapolate the data with the functional form
suggested by LO Wilson ChPT
\be
t_1^2 \chi_{\rm QCD} = c t_1 M^2_\pi + b \frac{a^2}{t_1}\; , 
\ee
where $t_1$ is a reference scale defined analogously to $t_0$. The function fits the data
well within the (large) statistical errors, see left plot in Figure~\ref{fig:chiQCDfull}.
By taking at face value the result of the fit of $c=2.8(5)\cdot 10^{-3}$ \cite{Bruno:2014ova},
by remembering that at LO in ChPT $c=F^2 t_1/2 N_f$, and by taking the value of
$F$ from Eq.~(\ref{eq:finalMeV}) and $t_1=0.061$~fm$^2$ from Ref.~\cite{Bruno:2014ova},
I get $N_f=2.06\pm0.38$. A result which is in agreement with the expected value, 
the error being still quite large though.

\section{Conclusions}
Over the last decade there has been an impressive global lattice
community effort to reach a precise quantitative understanding of
the behaviour of QCD in the chiral regime from first principles. A
rather clear and precise picture is emerging.

The spectral density of the Dirac operator in QCD Lite is non-zero at the origin
in the continuum and chiral limits. Its value coincide with $M_\pi^2 F_\pi^2/2m$
when $m\rightarrow 0$. This provides
a beautiful numerical proof that our picture of spontaneously broken
chiral symmetry in QCD is indeed correct. The low modes of the Dirac operator
do condense near the origin as dictated by the Banks-Casher mechanism, and
their rate of condensation generates the bulk of the pion mass 
up to quark masses that are about one order of magnitude larger than
in Nature.

The topological susceptibility shows the expected suppression with the mass
of the light quarks.
Within the (so far) large errors, results are compatible with LO ChPT. By now there
are many determinations of the QCD low-energy constants obtained by comparing the
predictions of ChPT with the precise lattice results with $N_f=2$,  $N_f=2+1$ and
$N_f=2+1+1$ flavours.

We have accumulated stronger and stronger evidence that the
breaking of the abelian chiral group due to the quantum anomaly is driven by the
Witten--Veneziano mechanism. This year it was understood that the recently found
definition of the topological charge via the Yang--Mills gradient-flow leads to cumulants
of the topological charge distribution which coincide with the universal
ones. All numerical results for the topological susceptibility in the SU($3$)
Yang--Mills theory obtained with the various (proper) definitions agree in the
continuum limit within the few percent precision. Its value reproduces the
mass of the $\eta'$ meson within the expected uncertainty.

On a more theoretical side, the (30 years) long story of defining the topological charge
on the lattice comes to a satisfactory end with a simple and elegant solution. Numerical
results are consistent with this theoretical progress.

Thanks to the conceptual, theoretical and technical advances achieved over the last decade
in lattice gauge theory, our femtoscope can explore the chiral regime
of QCD with higher and higher precision. This was just a dream only $10$-$15$ years ago, and
now is a reality. Hard work is still needed to empower the femtoscope in many
other corners of the theory, e.g. the baryon sector.

\section{Acknowledegments}
I wish to thank M. C\`e, C. Consonni, G. P. Engel, S. Lottini, M. L\"uscher
and R. Sommer for an enjoyable collaboration and for many inspiring discussions
during the last few years on the topics covered in this talk. I am grateful to
my colleagues within the CLS initiative and to those in the Alpha collaboration
for generating and sharing the ensembles
of gauge configurations with which many of the results reviewed here were
obtained. Many thanks to
H. Fukaya, K. Jansen, D.~J.~Murphy and S.~Schaefer for sending their results
before the conference, and for interesting discussions about their work.
Thanks to R. Sommer for a careful reading of this manuscript, and for many
suggestions to improve it. It is a pleasure to thank the organizers for their
great work in organizing the conference, for being able to create a very
stimulating atmosphere, and for giving me the honor to open the conference
with this talk.

\end{document}